\documentclass{article}
\usepackage{spconf,amsmath,graphicx}
\usepackage{enumitem}
\setlist{nosep, leftmargin=14pt}
\usepackage{mwe} 
\usepackage{hyperref}

\title{An unobtrusive quality supervision approach for medical image annotation}
\name{Sonja Kunzmann, Mathias \"Ottl, Prathmesh Madhu, Felix Denzinger, Andreas Maier}
\address{FAU Erlangen-N\"urnberg, \\
        Pattern Recognition Lab, \\ 
        Erlangen, Germany \\}

\begin{document}
%\ninept
%
\maketitle
\begin{abstract}
Image annotation is one essential prior step to enable data-driven algorithms. In medical imaging, having large and reliably annotated data sets is crucial to recognize various diseases robustly. However, annotator performance varies immensely, thus impacts model training. Therefore, often multiple annotators should be employed, which is however expensive and resource-intensive. Hence, it is desirable that users should annotate unseen data and have an automated system to unobtrusively rate their performance during this process. We examine such a system based on whole slide images (WSIs) showing lung fluid cells. We evaluate two methods the generation of synthetic individual cell images: conditional Generative Adversarial Networks and Diffusion Models (DM). For qualitative and quantitative evaluation, we conduct a user study to highlight the suitability of generated cells. Users could not detect 52.12\,\% of generated images by DM proofing the feasibility to replace the original cells with synthetic cells without being noticed.

\end{abstract}
%---------------------------------------
%---------------------------------------Introduction

\section{Introduction}
\label{sec:intro}
An essential part of Computer Vision is image annotation, which describes the enrichment of images with additional meta information of interest by a human reader. %labeling images and gives additional information to the image. 
It allows computers to gain a high-level understanding of images and to perceive the underlying information like humans. Labeling images is necessary for data sets because it lets a model know the essential parts of the image, i.e., classes of different types of cells. Especially in medical imaging, it is crucial to have trustful annotations. The annotations are the base for the machine to recognize various diseases, i.e., lung fluid cells describing asthma's insensitivity, which could further lead to cancer. However, ``to err is human'' and one big challenge of medical images is label noise, i.e. erroneously assigned labels. 
Annotation is a very time-consuming, subjective, and tedious task that is also expensive and often requires prior training. Primarily in the medical domain, i.e., for annotating cells in pathological whole slide images (WSIs), expert annotators usually have to get several years of practical experience. In other areas, crowdsourcing is a popular approach to obtaining label information. Crowdsourcing of annotations is performed by hosting data on an interactive platform like the EXACT online annotation tool \cite{Marzahl21}, where multiple annotators can assess the data simultaneously. Nevertheless, it is unknown whether the unseen data annotations can be trusted. 

We present our approach to unobtrusive measuring whether a person has correctly annotated the data and working on a concept to create semi-artificial images for previously unlabeled data. EXACT hosts a partially annotated WSI data set~\cite{Marzahl2021} for the cyto-pathological diagnosis of equine asthma, which we use in this approach. The goal is to have one fraction of real unlabeled images and one fraction of synthetic data where annotations are known. The differences of the data should not be recognizable or unobtrusive to the annotators. One step would be to generate a map of the WSI using image-to-layout and to replace the original lung fluid cells by labeling synthesized cells with inpainting. Additionally, we plan to use game elements like high scores to measure the accuracy of the annotator. Game elements count to gamification, representing the strategies taken from games, and are applied in a non-gaming context \cite{Morschheuser2018}. Gamification offers specific tools to increase motivation, i.e., with a high score on how well the annotator does the annotation process and creates higher performance expectations. Our concept would allow creating scores even for new unseen images, which could be used to motivate annotators, although it could potentially introduce bias in the annotations. 

%This paper considers the first step towards this concept by generating synthesized lung fluid cell images conditioned with a label using conditional Generative Adversarial Network (cGAN) and Diffusion Model (DM) (Fig. \ref{architecture}). We evaluated the synthetic images qualitatively with a user study in which people had to decide rather the images were synthesized (fake) or real. Only real vs. fake was chosen for a wider range, not a query on the lung fluid cell classes. %The user study's result gives information on our approach's other process.

Our contributions to progress in research include:
\begin{itemize}
    \item Proposal of a concept for the assessment of annotator performance on the task of WSI annotation.
    \item Creation of a set of simulated lung fluid cell images to enrich the abovementioned concept using two commonly generative approaches: cGAN and DM.
    \item Evaluation of the performance of these generative approaches in the scope of a user study.
\end{itemize}

\begin{figure*}[htb]
    \centering
    \includegraphics[width=0.6\textwidth]{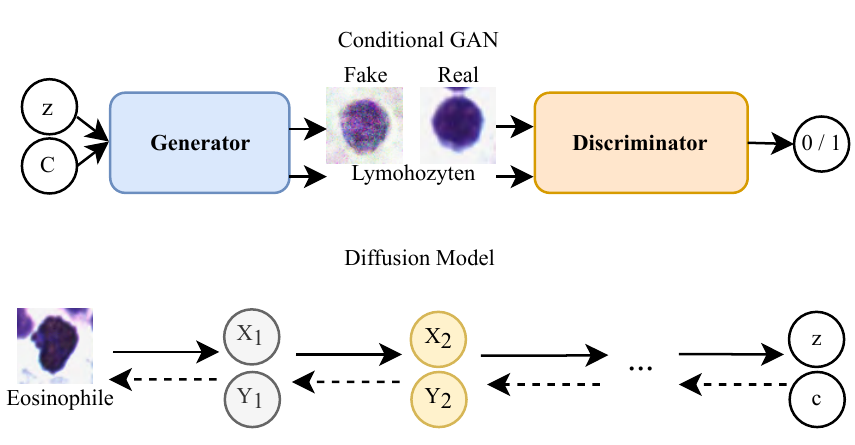}
    \caption{Architectures of the cGAN and Diffusion Model}
    \label{architecture}
\end{figure*}

%--------------------------------------- Material and Methods

\section{Material and Methods}
\label{sec:methods}
We present the data set characteristics used in this study and the architectures used to generate the lung fluid cells using cGAN~\cite{Mirza2014} and diffusion model~\cite{Rombach2021}. 

\subsection{Asthma Equidae Data Set}
\label{ssec:dataset}
\begin{figure*}[t]

    \begin{minipage}[b]{.48\textwidth}
        \centering
        \centerline{\includegraphics[width=0.8\textwidth]{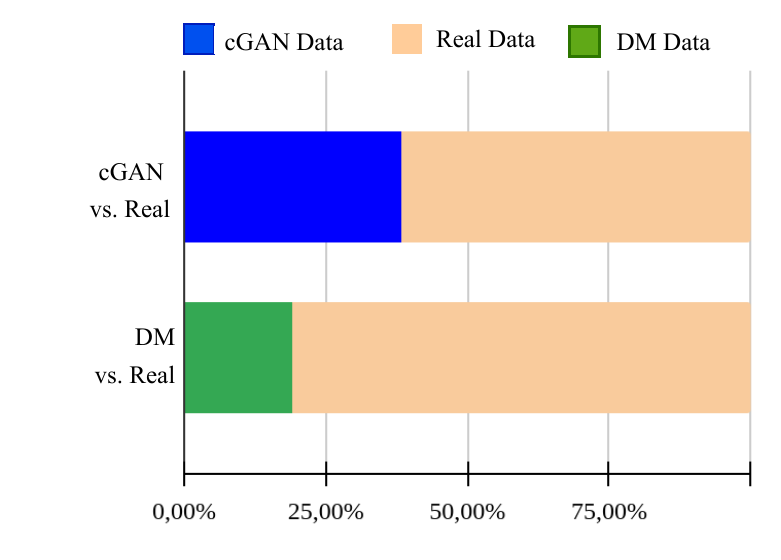}}
    %   \vspace{1.5cm}
        \caption{Pick rate of our user study with represented the synthesized image over real lung fluid cells}
        \label{pick_rate}
    \end{minipage}
    \hfill
    \begin{minipage}[b]{0.48\textwidth}
        \centering
        \centerline{\includegraphics[width=0.8\textwidth]{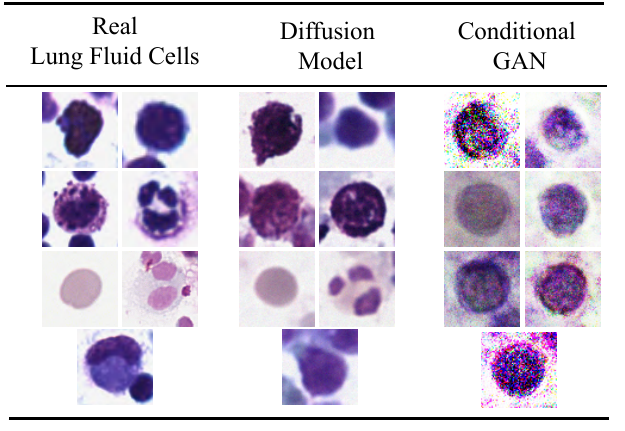}}
    %   \vspace{1.5cm}
        \caption{Visualization of a few examples of the original and generated synthetic lung fluid cells}
        \label{results}
    \end{minipage}

\end{figure*}

The Asthma Equidae data set includes six cytological samples of equine bronchoalveolar lavage fluid (BALF), wherein the single cells are cropped from WSIs. This data set was supplied by Marzahl et al.~\cite{Marzahl2021}. WSIs were cytocentrifugated and shaded using the May Grunwald Giemsa stain. Cytocentrifuge describes concentrating cells in liquid samples on a WSI to be stained and examined. Afterward, the glass slides were digitized with a linear scanner of Aperio ScanScope CS2, Leica Biosystems in Germany at a magniﬁcation of 400x with a resolution of 0.25~$\mu$m/px~\cite{Marzahl2021}. A trained pathologist partially annotated two of the six WSIs fully and four partially. The $87{,}738$ bounding box annotated cells include neutrophils ($12{,}556$), multinuclear cells ($310$), mast cells ($1{,}553$), macrophages ($24{,}498$), lymphocytes ($46{,}397$), erythrocytes ($339$) and eosinophils ($105$)~\cite{Marzahl2021}. A few randomly chosen examples of the individual lung fluid cell types can be seen in the left-most column of Figure~\ref{results}. 

\subsection{Conditional Generative Adversarial Network}
\label{ssec:cGAN}
Generative Adversarial Networks (GAN)~\cite{GAN} are an adversarial model comprising of a generative and a discriminative model, shown at the top of Fig. ~\ref{architecture}.
The Generator replicates the data distribution, and the discriminator calculates the probability of belonging to real over synthesized samples from the training data ~\cite{Mirza2014}. It constitutes a two-player min-max game with a value function. However, cGANs modify the original GAN model. In a cGAN, the discriminator is given data-label pairs (i.e., input or output) instead of just data. The Generator is given a label in addition to the noise vector, indicating which data classes the image should belong to ~\cite{Mirza2014}. The addition of labels forces the Generator to learn multiple representations of different training data classes, allowing the ability to explicitly control the generator's output ~\cite{Mirza2014}. The generator tries to imitate the real data distribution by minimizing the loss between the real and synthesized data, while the discriminator tries to maximize the probability of discriminating between real and synthesized data. Therefore, this process is also called a two-player min-max game. 

\subsection{Diffusion Model}
\label{ssec:DM}
A DM has two significant domains of processes~\cite{Rombach2021}. In the forward stage, the image is corrupted by gradually introducing noise until the image becomes completely random noise. In the reverse process, a U-net gradually removes the predicted noise at each time step. Any generative learning method has two main stages: perceptual compression and semantic compression ~\cite{Rombach2021}. During a perceptual compression learning phase, a learning method must encapsulate the data into an abstract representation by removing the high-frequency details. This step is necessary for building a translation invariant and robust conceptual representation of an environment. In the semantic phase of learning, an image generation method must be able to capture the semantic structure present in the data. This conceptual and semantic structure preserves the context and inter-relationship of various objects in the image.

%\subsection{Turing Test}
%\label{ssec:turning}
%The Turing test is a method for determining whether a machine can think the same way as a human. It was developed in 1950 by Alan Turing \cite{Turing1950}. Turing proposed that a human rater should evaluate natural language conversations between humans and machines designed to produce human-like responses. The rater would know that one of the two players in the talk was a machine. The conversation would be limited to a computer keyboard and screen, so the outcome would not depend on the machine's capacity to effect words as speech. The test is passed if the tester cannot tell which of the participants is the human and who the machine is. The test result would not depend on the machine's ability to answer correctly but on how closely its answers resemble a human's \cite{Turing1950}.

%--------------------------------------- Experiment and Results

\section{Experiments and Results}
\label{sec:exp}
This section presents the experimental setup and results for the proposed approach.
% In this section, we go into detail about the experiment design and the results of our approach to think about further steps. 
\subsection{Experimental Setup}
\label{ssec:design}
First, we generate lung fluid cells by using cGAN and DM. The cGAN is trained for 200 epochs with an input resolution of $64\times64$ with the following parameters: three channels, a batch size of 64, a learning rate of 0.0002, an Adam optimizer, and the binary cross entropy (BCE) loss with logits loss which is a combination of BCE and a Sigmoid layer. The data set is balanced using class weights within the loss function to tackle the severe class imbalance.  The network was configured with five alternating layers of linear and LeakyReLU. The model generated 1050 samples per class with the above parameters and was chosen randomly from the data set. The DM architecture design and implementation is based on the publicly available on GitHub repository (available at \url{https://github.com/CompVis/latent-diffusion}), but we used only the diffusion model without the autoencoder part. 

The DM was trained until the loss stagnated, in our case, 75 epochs. Lung fluid cell classes were randomly oversampled to tackle the severe class imbalance. The data set of seven classes of lung fluid cell types were divided into training (80\,\%), testing (10\,\%), and validation (10\,\%) sets. 
Next, we conducted a user study with a Turing Test to estimate if humans can distinguish synthesized (fake) from real data. Random samples were taken from the original lung fluid cells data set and the synthesized cells of cGAN and DM results. The study was divided into two parts, and we imposed no time limit for labeling. All samples were resized to $64\times64$. 

In the first sub-study, all participants had to decide between one real and one synthesized sample of 20 pairs. The pairs were separated into ten pairs showing a synthesized image from the cGAN, and a real sample and ten synthesized images from the DM and a real sample. The participant had to choose the fake sample. The second sub-study included 30 questions showing only a single sample for which the participant had to decide whether it was real or fake. The participants could randomly see ten cGAN, ten DM, and ten real lung fluid cell images. To better evaluate the results, we decided to use the pick rate. The pick rate is the percentage of one image chosen over its counterpart based on the total number of choices. The confusion matrix is used to calculate Accuracy, Recall, and Precision.  

\begin{figure}[t]
    \begin{minipage}[b]{1.0\linewidth}
      \centering
          \centerline{\includegraphics[width=0.6\linewidth]{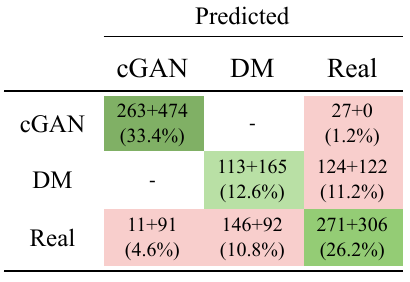}}
          \caption{Results from our user study, confusion matrix (absolute and relative values)}
          \label{cf}
    \end{minipage}
\end{figure}

\subsection{Results and Discussion}
\label{ssec:results}
We presented two methods suitable for generating synthetic lung fluid cells (see Fig. ~\ref{architecture}). Fig. \ref{results} shows the results from left to right, the first row of real data, the second row of DM results, and the last row of cGAN samples. The next step was to execute the user study with randomly chosen samples of cGAN, DM, and real data. In total, 49 people enrolled in our study with different background knowledge in the field of computer science (53.2\,\%), medical engineering (23.4\,\%), and others (23.4\,\%). Of those people, 10.4 \,\% saw lung fluid cells before. Most participants took six to eight minutes to answer the user study. We took this information not into account for this study. We calculate the pick rate to get a better overview (Fig. ~\ref{pick_rate}).

The pick rate of cGAN (38.25\,\%) is much higher than the pick rate of DM (18.94\,\%), which means that the cGAN results were more often chosen as fake than DM results because they were more challenging to recognize. One reasonable explanation for the higher pick rate of cGAN results could be the noise in these samples (see Fig. ~\ref{results}). 

Additionally, we present a confusion matrix (Fig. ~\ref{cf}) that shows that 46.0\,\% were recognized as fake images. The confusion matrix shows a combination of parts one and two. The accuracy of cGAN results is 91.30,\% and for DM results is 47.88\,\%, which means that cGAN results are more easily recognized as DM samples in part one. As we aim to produce images that will be confused with real images, the annotator needs to notice that the WSI includes synthetic cells with ground truth, leading to low accuracy. The precision shows that 95.99\,\% of cGAN samples were predictive compared to DM, which is at 43.63\,\%. The recall with 90.69\,\% confirms our assumption and shows that the cGAN outputs have almost been correctly titled as "fake". In contrast, the DM results have a recall of 47.68\,\%.  

The user study shows that the DM results look more similar to the real lung fluid cells. So the DM is better able to produce realistic-looking lung fluid cell images than cGANs. With a pick rate of 18.94\,\% for DM generated fake cell images, they resemble real lung fluid cell images sufficiently. Consequently, it is harder to distinguish them, but there are still further challenges, like the background of a lung fluid cell, different image sizes, and color intensity. The cGAN images do not look optimal, which indicates instability of the training, which is a known problem and leads to more impaired quality. We need additional synthetic data, especially of classes that do not have enough ground truth, i.e., multinuclear cells ($310$). 

%--------------------------------------- Conclusion

\section{Conclusion}
\label{sec:conclusions}
Image annotation in medical image data sets is essential to the deep learning pipeline because they recognize various diseases strenuously. Hence, the labeling process is very relevant to get the valid ground truth. Our idea is that annotators should label unseen data, and an automated system unobtrusively rates their performance during this process. We followed the pipeline of our contributions by generating synthesized lung fluid cell images conditioned with a label using a conditional Generative Adversarial Network (cGAN) and Diffusion Model (DM). Then conducted a user study to determine if annotators see a big difference between real and generated synthetic lung fluid cells, with the result that the cGAN images could easily be identified while the DM tricked most human readers.

In future work, we will address the remaining steps of a quality and unobtrusive supervision process for annotators. First, we plan to use an image-to-layout synthesis architecture to create a map of the whole WSI so that one-half of the real lung fluid cells can be exchanged with synthetic cells. A problem that could arise is that in WSI, fake cells might be recognized by their differing background and size. Our presented user study did not cover this possibility and should be addressed in future works. Additionally, we plan to use an inpainting method to insert desired lung fluid cell image. An essential part is that the annotator is unobtrusive of the synthetic data and has to annotate the entire WSI. 
Secondly, based on the known class label for the synthesized lung fluid cells, a high score can be calculated for how accurately the annotator has annotated using game elements.

\section{Compliance with ethical standards}
\label{sec:ethics}
All procedures performed in studies involving animal subjects were under the ethical standards of the institutional and/or national research committee and with the 1964 Helsinki declaration and its later amendments or comparable ethical standards. %CAB gratefully acknowledges the financial support received from the Dres. Jutta & Georg Bruns-Stiftung für innovative Veterinärmedizin.

\section{Acknowledgments}
\label{sec:acknowledgments}
The research leading to these results has received funding from the European Research Council (ERC) under the European Union’s Horizon 2020 research and innovation program (ERC Grant no. 810316). The hardware is funded by the German Research Foundation (DFG). Felix Denzinger is an employee of Siemens Healthcare GmbHF, and the other authors declare no conflicts of interest. 

\bibliographystyle{IEEEbib}
\bibliography{refs}

\end{document}